\documentclass[Physsubmission, Phys]{SciPost}

\hypersetup{
    colorlinks,
    linkcolor={red!50!black},
    citecolor={blue!50!black},
    urlcolor={blue!80!black}
}

\usepackage[bitstream-charter]{mathdesign}

\DeclareSymbolFont{usualmathcal}{OMS}{cmsy}{m}{n}
\DeclareSymbolFontAlphabet{\mathcal}{usualmathcal}

\begin{document}

\begin{center}{\Large \textbf{
What is the mechanism of the $T_{4c}(6900)$ tetraquark production ?
}}\end{center}

\begin{center}
A. Szczurek \textsuperscript{1,2},
R. Maciula \textsuperscript{1} and
W. Sch\"afer \textsuperscript{1$\star$}
\end{center}

\begin{center}
{\bf 1} Institute of Nuclear
Physics, Polish Academy of Sciences, ul. Radzikowskiego 152, 
PL-31-342 Krak{\'o}w, Poland
\\
{\bf 2} College of Natural Sciences, Institute of Physics,
University of Rzesz\'ow, ul. Pigonia 1, PL-35-959 Rzesz\'ow, Poland
\\
* antoni.szczurek@ifj.edu.pl
\end{center}

\begin{center}
\today
\end{center}


\definecolor{palegray}{gray}{0.95}
\begin{center}
\colorbox{palegray}{
  \begin{tabular}{rr}
  \begin{minipage}{0.1\textwidth}
    \includegraphics[width=22mm]{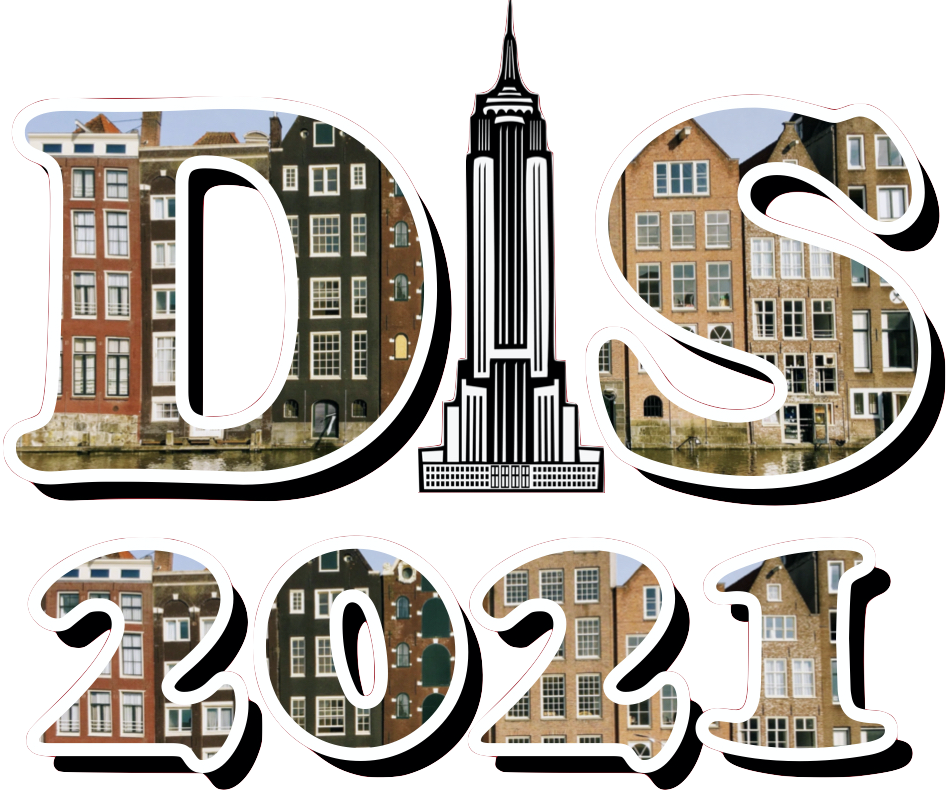}
  \end{minipage}
  &
  \begin{minipage}{0.75\textwidth}
    \begin{center}
    {\it Proceedings for the XXVIII International Workshop\\ on Deep-Inelastic Scattering and
Related Subjects,}\\
    {\it Stony Brook University, New York, USA, 12-16 April 2021} \\
    \doi{10.21468/SciPostPhysProc.?}\\
    \end{center}
  \end{minipage}
\end{tabular}
}
\end{center}

\section*{Abstract}
{\bf
We discuss the production mechanism of fully charm tetraquark, observed
recently by the LHCb at M = 6.9 GeV in the $J/\psi J/\psi$ channel.
Both single parton scattering (SPS) and double parton scattering (DPS)
mechanisms are considered.
We calculate the distribution in the invariant 
mass of the four-quark system $M_{4c}$ for SPS and DPS production of 
$c c \bar c \bar c$ in the $k_t$-factorization approach.
We present transverse momentum distribution in $p_{t,4c}$
for $c \bar c c \bar c$ system in a mass window around tetraquark mass.
The calculation of the SPS
$g^* g^* \to T_{4c}(6900)$ fusion mechanism is performed in the
$k_T$-factorization approach assuming different spin scenarios ($0^+$
and $0^-$).
}

\vspace{10pt}
\vspace{10pt}

\section{Introduction}
\label{sec:intro}

The recent observation by the LHCb collaboration \cite{LHCb_T4c} 
of a sharp peak in the di-$J/\psi$ channel at $M$ = 6.9 GeV seems 
to strongly suggest the presence of a fully charm tetraquark, 
consisting of $c c \bar c \bar c$.
A number of theoretical models for the spectroscopy of 
tetraquarks were developed.
The most popular approach treats the fully heavy ($c c \bar c \bar c$,
$b b \bar b \bar b$ or $c \bar c b \bar b$) tetraquarks as a bound system
of a color antitriplet diquark and color triplet antidiquark.

Assuming the peak observed by LHCb is indeed caused by a new
state, models suggest that it is rather an excited state.
However, the spin and parity of the state are not known at present.

Different $J^{PC}$ combinations are possible in general
\cite{DN2019,BFRS2019,CCLZ2020,LCD2020},
What is the mechanism of the fusion of four charm
quarks/antiquarks is not clear in the moment. 
A large cross section for 
$c \bar c c \bar c$ production at the LHC due to
double-parton scattering (DPS) mechanism was predicted in
\cite{LMS2012}.
The contribution of the single parton scattering 
to the $c \bar c c \bar c$ production 
is much smaller \cite{SS2012,MS2013,HMS2014,HMS2015}.
The production of two $J/\psi$ states was also studied at the
LHC \cite{LHCb_jpsijpsi,CMS_jpsijpsi,ATLAS_jpsijpsi,Baranov:2012re}.

Recently, our group studied production of pseudoscalar 
\cite{babiarz_pseudoscalar} and scalar \cite{babiarz_scalar} charmonia
via gluon-gluon fusion. We shall use the same formalism also
in the context of the tetraquark production.

\section{Basic formalism}

\begin{figure}
\begin{center}
\includegraphics[width=4.5cm]{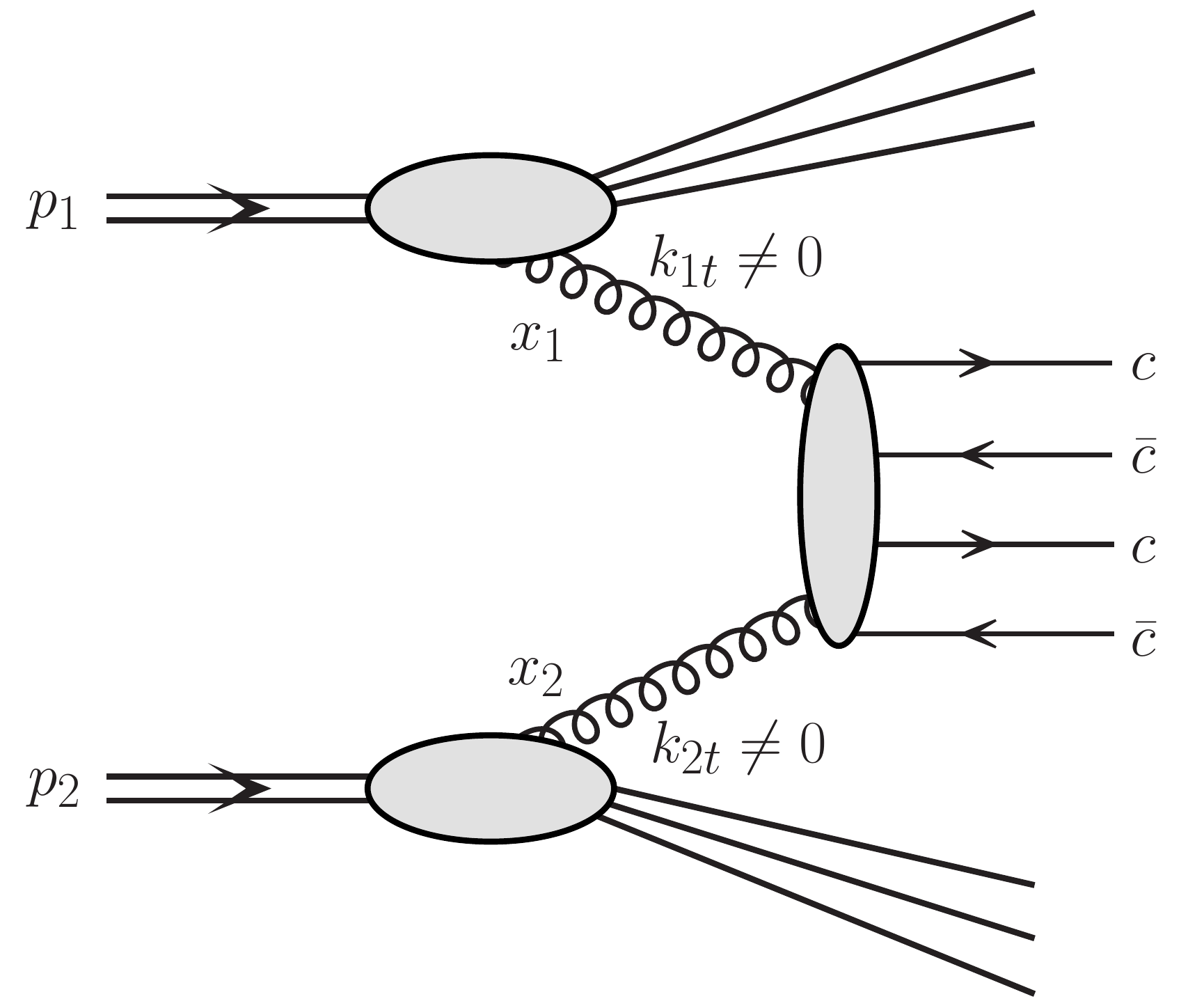}
\includegraphics[width=4.5cm]{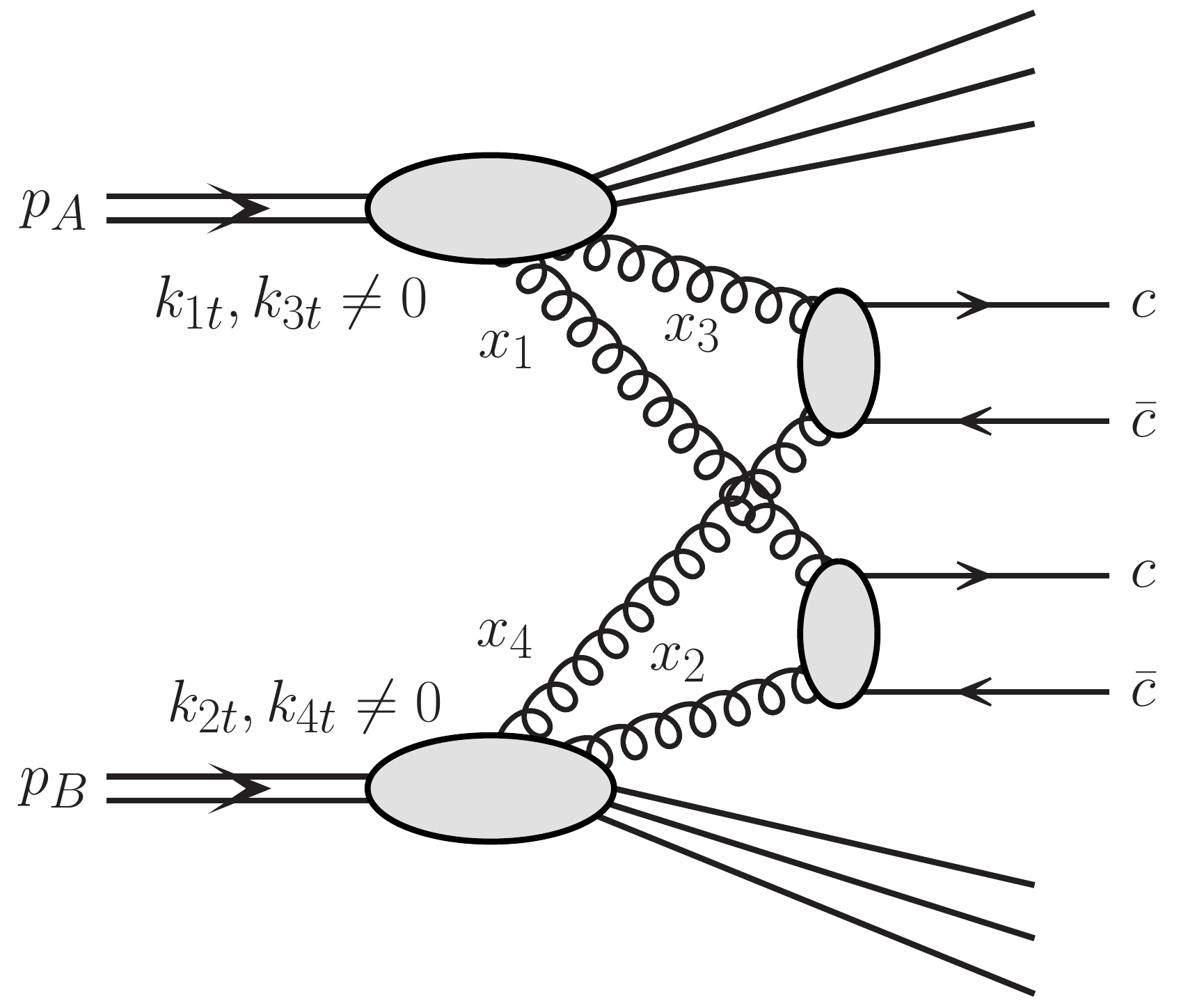}
\caption{Two dominant reaction mechanisms of production of 
$c \bar c c \bar c$ nonresonant continuum. The left diagram represent
the SPS mechanism (box type) and the left diagram the DPS mechanism.
}
\label{fig:diagrams_ccbarccbar}
\end{center}
\end{figure}

In Fig.\ref{fig:diagrams_ccbarccbar} we show the dominant reaction
mechanisms of $c c \bar c \bar c$ production: 
SPS type (left diagram) and DPS type (right diagram).

In our recent study both the SPS and the DPS contributions are
calculated in the framework of $k_{T}$-factorization \cite{MSS2021}.

In this approach the SPS cross section for 
$pp \to c\bar c c\bar c \, X$ reaction can be written as
\begin{equation}
d \sigma_{p p \to c\bar c c\bar c \; X} =
\int d x_1 \frac{d^2 k_{1t}}{\pi} d x_2 \frac{d^2 k_{2t}}{\pi}
{\cal F}_{g}(x_1,k_{1t}^2,\mu^2) {\cal F}_{g}(x_2,k_{2t}^2,\mu^2)
d {\hat \sigma}_{g^*g^* \to c\bar c c\bar c}
\; .
\label{cs_formula}
\end{equation}
Above ${\cal F}_{g}(x,k_t^2,\mu^2)$ is the unintegrated
gluon distribution function (gluon uPDF).

The elementary cross section in Eq.~(\ref{cs_formula}) can be written as:
\begin{equation}
d {\hat \sigma}_{g^*g^* \to c \bar c c \bar c} = { 1 \over (2!)^2}
\prod_{l=1}^{4}
\frac{d^3 \vec p_l}{(2 \pi)^3 2 E_l} 
(2 \pi)^4 \delta^{4}(\sum_{l=1}^{4} p_l - k_1 - k_2) \frac{1}{\mathrm{flux}} \overline{|{\cal M}_{g^* g^* \to c \bar c c \bar c}(k_{1},k_{2},\{p_l\})|^2}
\; .
\label{elementary_cs}
\end{equation}

\begin{figure}
\begin{center}
\includegraphics[width=4.5cm]{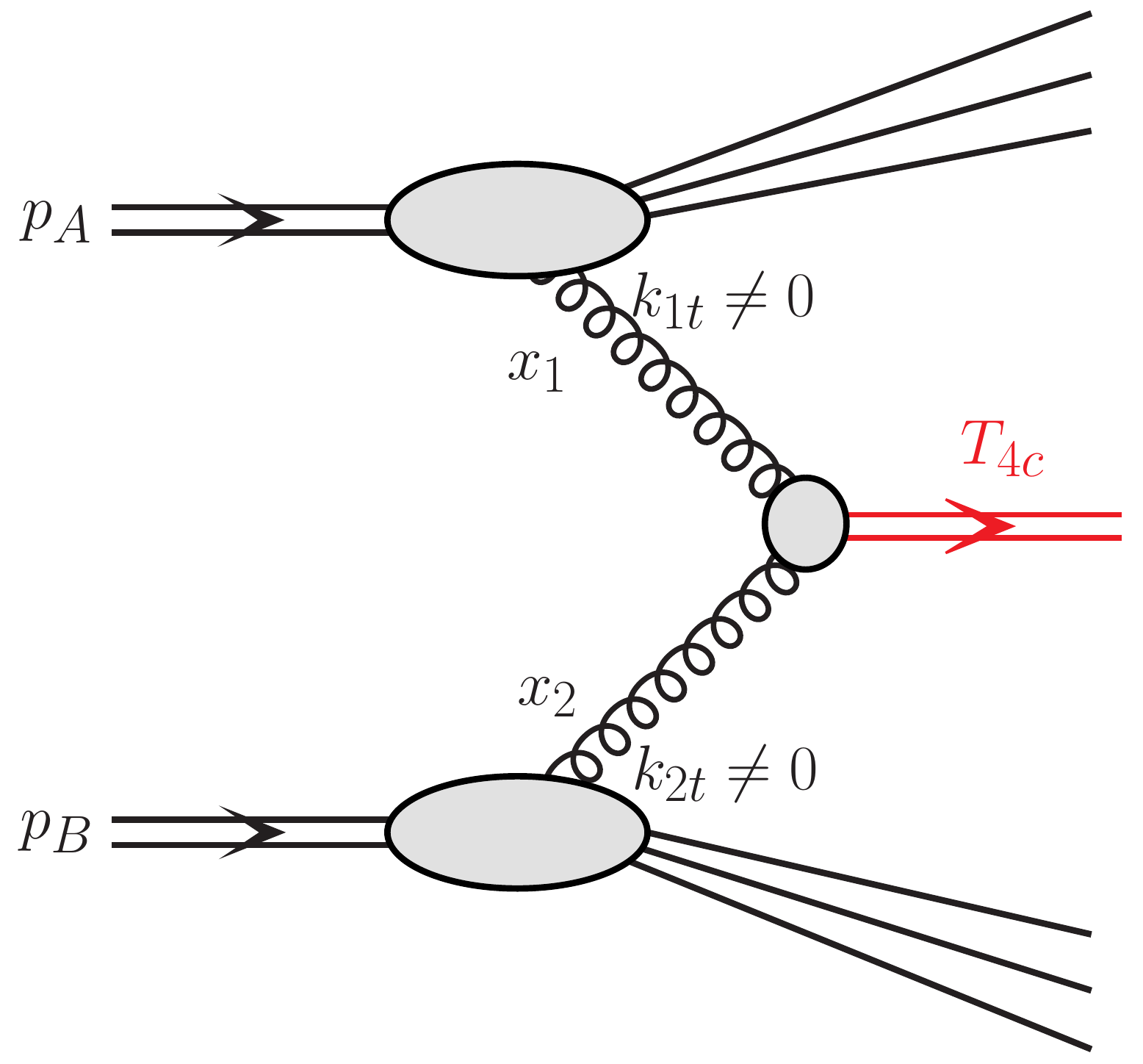}
\caption{The mechanism of gluon-gluon fusion leading to the production
of the $T_{4c}(6900)$ tetraquark.
}
\label{fig:pp_T4c}
\end{center}
\end{figure}

We consider also the calculation of the SPS-type signal
as a fusion of two (off-shell) gluons for two different spin-parity
assignments: $0^+$ and $0^-$ of the tetraquark. The corresponding
diagram is shown in Fig.\ref{fig:pp_T4c}.

In this context we use the formalism worked out recently for 
the inclusive production of pseudoscalar \cite{babiarz_pseudoscalar} 
and scalar \cite{babiarz_scalar} quarkonia. 
The off-shell gluon fusion cross sections is proportional to a
form-factor, which depends on the virtualities of gluons, $Q_i^2 = - k_i^2$:
\begin{eqnarray}
d \sigma_{g^* g^* \to 0^-} &\propto& {1 \over k_{1t}^2 k_{2t}^2} \,
(\vec k_{1t} \times \vec k_{2t})^2 \, F^2(Q_1^2,Q_2^2) \, , \nonumber \\
d \sigma_{g^* g^* \to 0^+} &\propto&  {1 \over k_{1t}^2 k_{2t}^2}  \Big( (\vec k_{1t} \cdot \vec k_{2t}) (M^2 + Q_1^2 + Q_2^2) + 2 Q_1^2 Q_2^2 \Big)^2 \,  {F^2(Q_1^2,Q_2^2) \over 4X^2} \, ,
\end{eqnarray} 
with $X = (M^4 + 2(Q_1^2+Q_2^2)M^2+(Q_1^2-Q_2^2)^2)/4$.
Note, that for the $0^+$ assignment we use only the TT coupling, as in analogy with \cite{babiarz_scalar} 
we expect the LL contribution to be smaller.

\begin{figure}
\begin{center}
\includegraphics[width=4.5cm]{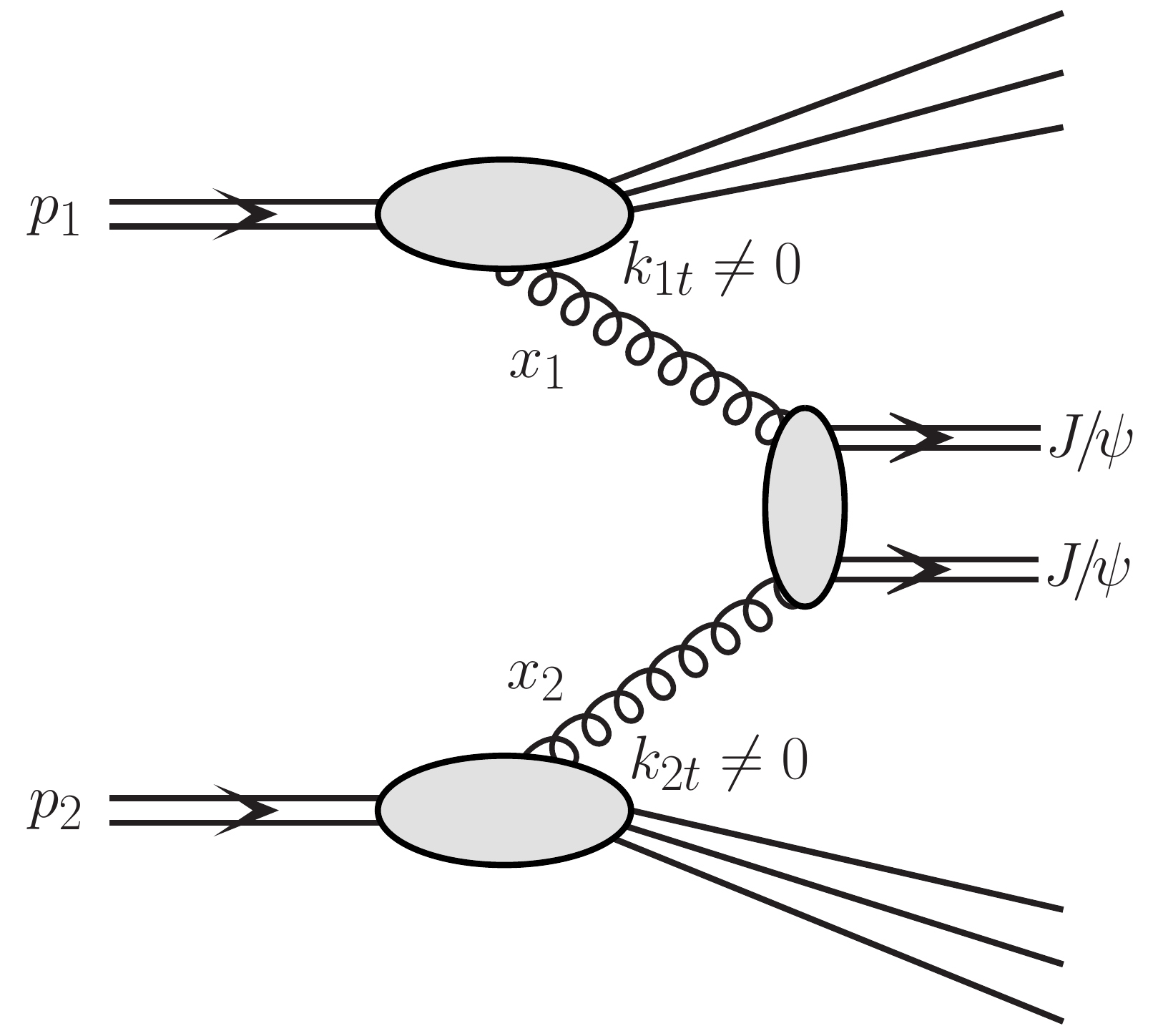}
\includegraphics[width=4.5cm]{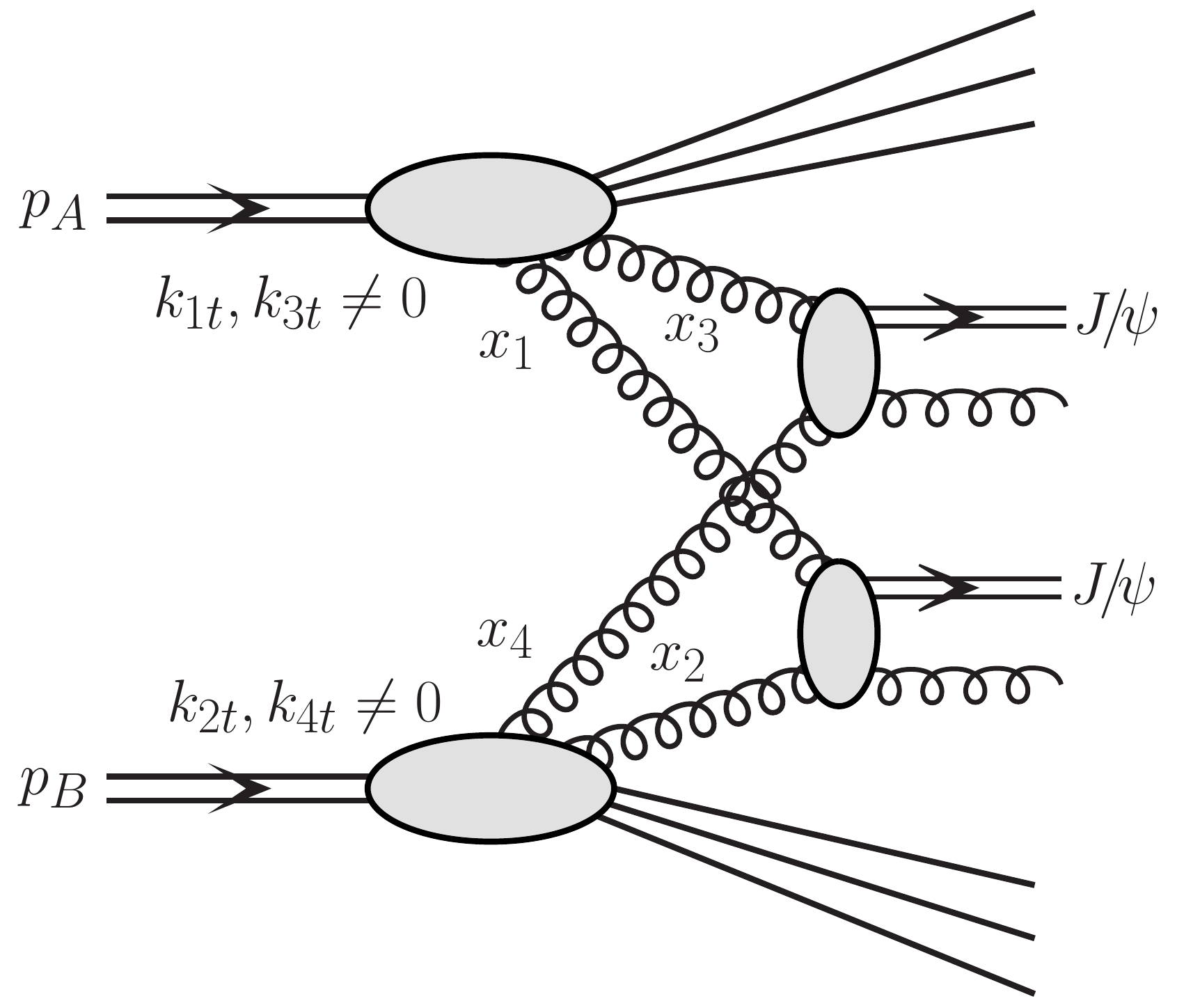}
\caption{Two dominant reaction mechanisms of production of 
$J/\psi J/\psi$ nonresonant continuum. The left diagram represent
the SPS mechanism (box type) and the right diagram the DPS mechanism.}
\label{fig:diagrams_jpsijpsi}
\end{center}
\end{figure}

The mechanisms of the background productions are shown in
Fig.\ref{fig:diagrams_jpsijpsi}.

\section{Selected results}
\label{sec:results}

\begin{figure}
\begin{center}
\includegraphics[width=.45\textwidth]{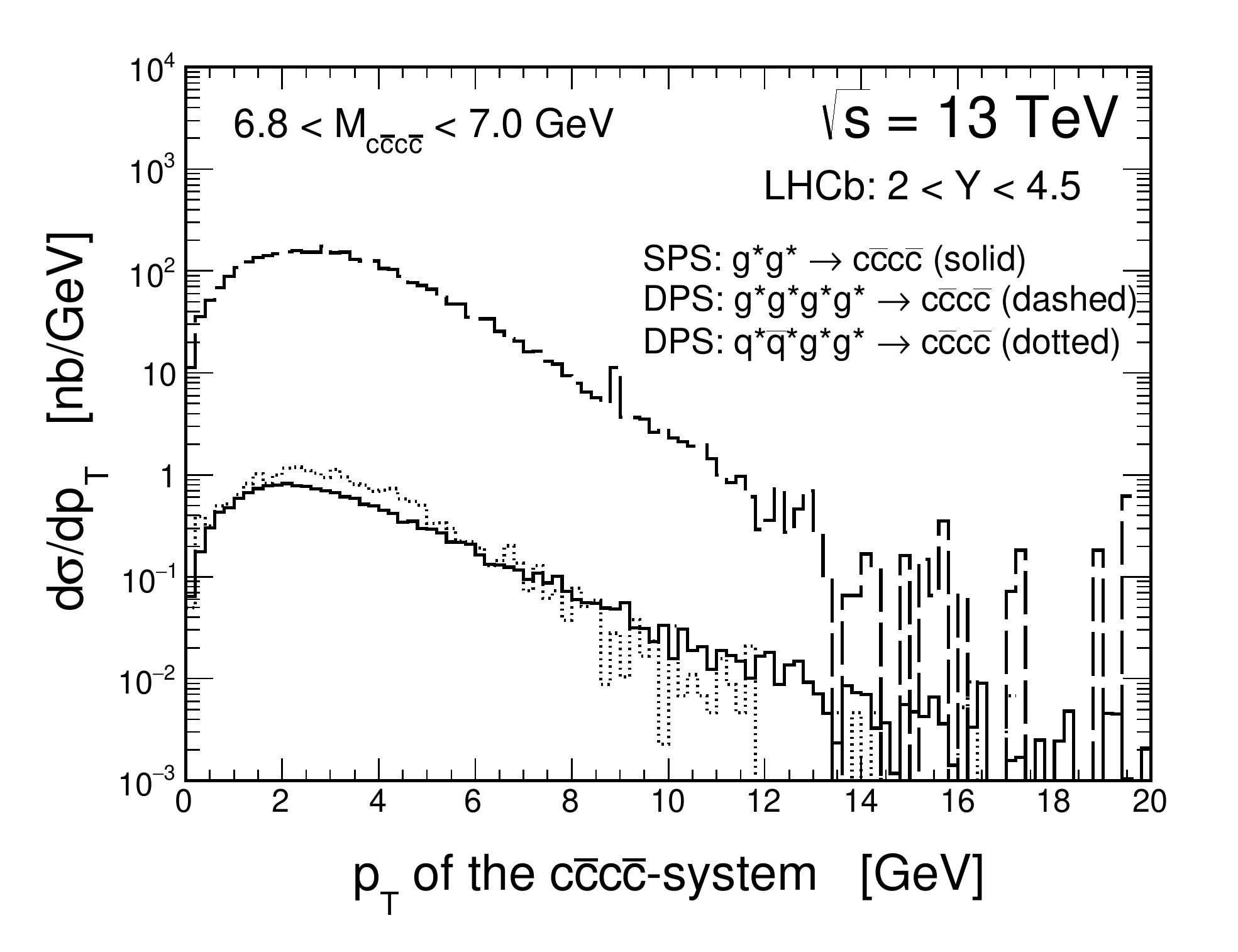}
\caption{Distribution of $p_{t,4c}$ of four quark-antiquark system
within invariant mass window $(M_R - \rm{0.1 GeV}, M_R + \rm{0.1 GeV})$.
Here $\sqrt{s} =$ 13 TeV and average rapidity of quarks and antiquarks
in the interval (2,4.5). The solid line is for SPS, the
dashed line for $g g g g \to c \bar c c \bar c$ DPS 
and the dotted line is for $q \bar q g g \to c \bar c c \bar c$ DPS 
contribution.
}
\label{fig:dsig_dpt_4c}
\end{center}
\end{figure}

In Fig.\ref{fig:dsig_dpt_4c} we present distribution in 
$p_{t,4c}$ of four quark-antiquark system
within invariant mass window $(M_R - \rm{0.1 GeV}, M_R + \rm{0.1 GeV})$.
In a naive coalescence model this corresponds to the distribution of 
the tetraquark.

\begin{figure}
\begin{center}
\includegraphics[width=.4\textwidth]{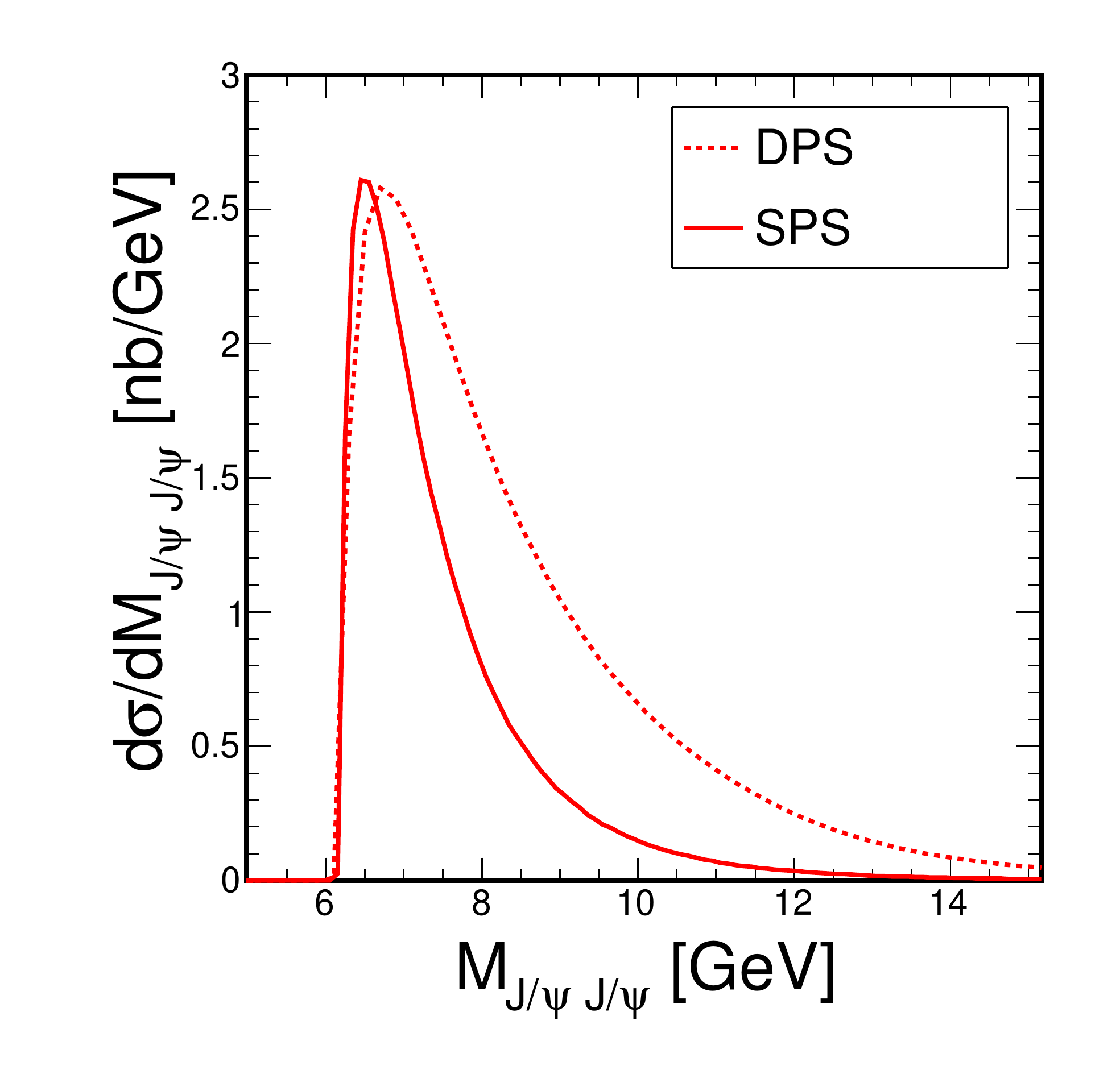}
\caption{Distribution in invariant mass of the $J/\psi J/\psi$ system
for SPS (solid line) and DPS (dashed line).
In this calculation $\sqrt{s}$ = 13 TeV and we assumed that both
$J/\psi$ mesons have rapidity in the (2,4.5) interval.}
\label{fig:dsig_dMVV}
\end{center}
\end{figure}

In Fig.\ref{fig:dsig_dMVV} we show distribution in $M_{J/\psi J/\psi}$
for the two background mechanisms shown in Fig.\ref{fig:diagrams_jpsijpsi}.
We see that in the vicinity of the tetraquark mass the SPS contribution
is similar to the DPS one so both of them must be 
included in the evaluation of the background.

\begin{figure}
\begin{center}
\includegraphics[width=.36\textwidth]{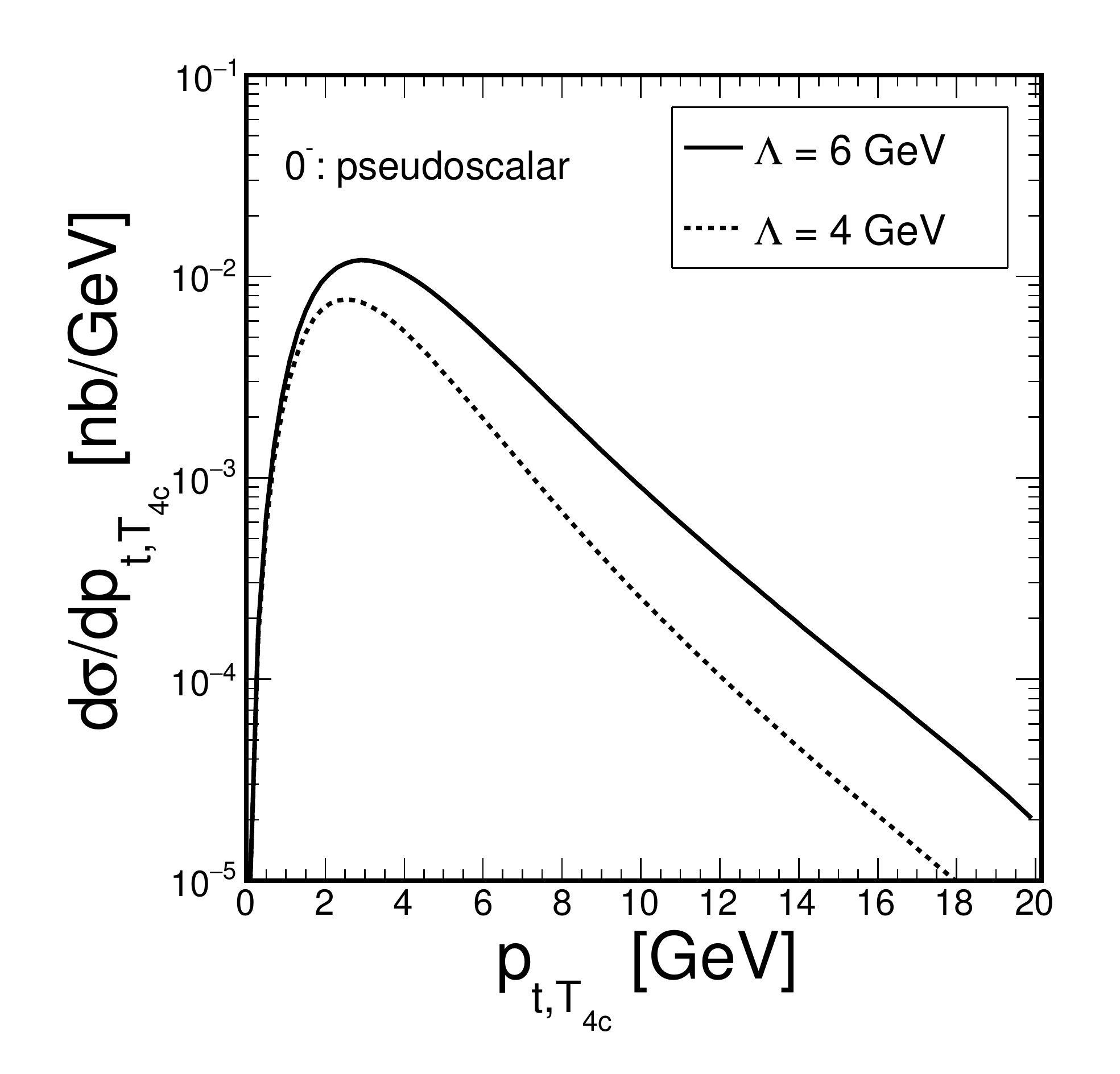}
\includegraphics[width=.36\textwidth]{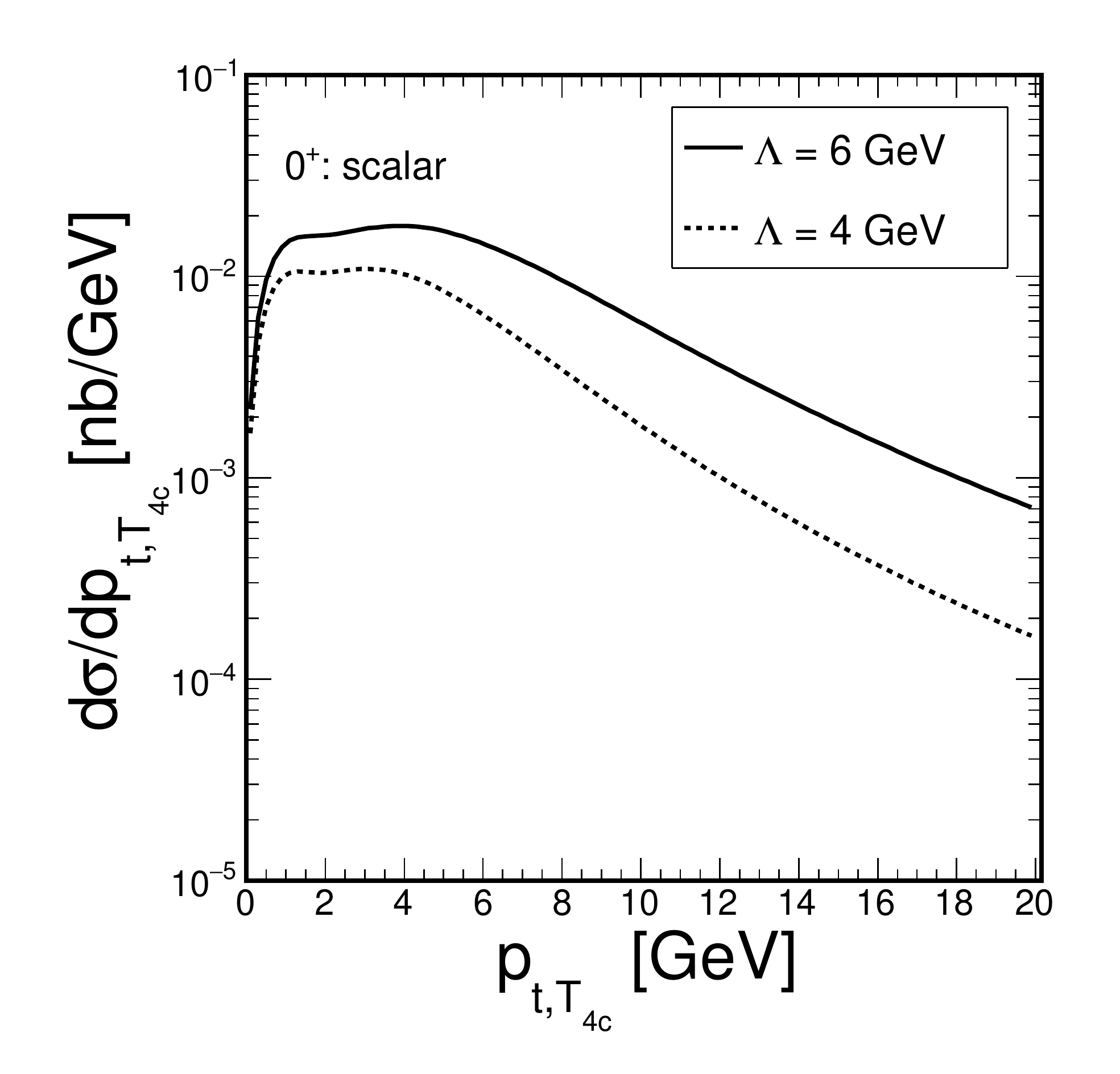}
\caption{Transverse momentum distribution of the $T_{4c}(6900)$
tetraquark for the $0^-$ (left panel) and $0^+$ (right panel)
assignments. Here $\sqrt{s}$ = 13 TeV. We show results for the
KMR UGDF and $\Lambda$ = 6 GeV (solid line) and 
$\Lambda$ = 4 GeV (dashed line).
}
\label{dsig_dpt_resonance}
\end{center}
\end{figure}

The transverse momentum distribution of the $T_{4c}(6900)$ tetraquark
produced in a single particle mechanism is shown in 
Fig.\ref{dsig_dpt_resonance}.

Since the ratio of signal-to-background improves with transverse
momentum of the tetraquark \cite{LHCb_T4c} and knowing
relatively well the behaviour of the SPS and DPS background 
\cite{MSS2021} we can conclude that 
the $0^-$ assignment is disfavoured by the LHCb experimental results. 

\section{Conclusion}

In our recent paper \cite{MSS2021} we have considered several 
aspects related to
the production of $T_{4c}(6900)$ tetraquark  observed
recently by the LHCb collaboration in the $J/\psi J/\psi$ channel and 
the $J/\psi J/\psi$ background. 

While the background distributions can in our opinion be reliably 
calculated, it is not the case for the signal. 
In the discussed naive coalescence model we have adjusted 
a normalization factor $C$ responsible for
the formation probability $P_{T_{4c}}$ and decay branching fraction 
$Br(T_{4c}(6900) \to J/\psi J/\psi)$.
We have obtained $C$ = 10$^{-4}$ for the DPS and C = 10$^{-2}$ for the SPS
production of $c \bar c c  \bar c$.

We have considered also more explicitly the SPS mechanism
of the resonance production via gluon-gluon fusion in 
the $k_T$-factorization approach. Also in this 
case the normalization, related to the underlying formation process
and/or wave function of the tetraquark
and the decay branching fraction $T_{4c} \to J/\psi J/\psi$ must 
be adjusted to the experimental signal-to-background ratio.
In this study we have considered two examples of the $0^+$ and $0^-$
assignment. The current data seem to exclude the $0^-$ assignment
as the final result contradicts qualitatively to the transverse 
momentum dependence of the signal-to-background ratio as observed 
by the LHCb collaboration \cite{LHCb_T4c}.

\section*{Acknowledgements}
This study was partially supported by the Polish National Science Center
grant UMO-2018/31/ /B/ST2/03537 and by the Center for Innovation and
Transfer of Natural Sciences and Engineering Knowledge in Rzesz{\'o}w.




\begin{thebibliography}{100}

\bibitem{LHCb_T4c}
R. Aaij et al. (LHCb collaboration),
``Observation of structure in the $J/\psi$-pair mass spectrum'',
Sci. Bull. \textbf{2020}, 65
[arXiv:2006.16957 [hep-ex]].

\bibitem{DN2019}
V.R. Debastiani and F.S. Navarra,
``A non-relativistic model for the $[c c][\bar c \bar c]$ tetraquark'',
Chin. Phys. {\bf C43}, 013105 (2019).

\bibitem{BFRS2019}
M.A. Bedolla, J. Ferretti, C.D. Roberts and E. Santopinto,
``Spectrum of fully-heavy tetraquarks from a diquark+antidiquark
perspective'',
Eur. Phys. J. C \textbf{80}, no.11, 1004 (2020)
[arXiv:1911.00960 [hep-ph]].

\bibitem{CCLZ2020}
H.X. Chen, W. Chen, X. Liu and S.-L. Zhu,
``Strong decays of fully-charm tetraquarks into di-charmonia'',
Sci. Bull. \textbf{65}, 1994-2000 (2020)
[arXiv:2006.16027 [hep-ph]].

\bibitem{LCD2020}
Q.-F. Lu, D.-Y. Chen and Y.B. Dong,
``Masses of fully heavy tetraquark $Q Q \bar Q \bar Q$ in an extended
relativized quark model'',
Q.~F.~L\"u, D.~Y.~Chen and Y.~B.~Dong,
Eur. Phys. J. C \textbf{80} (2020) no.9, 871
[arXiv:2006.14445 [hep-ph]].

\bibitem{LMS2012} 
M. {\L}uszczak, R. Maciu{\l}a and A. Szczurek,
``Production of two $c \bar c$ pairs in double-parton scattering'',
Phys. Rev. {\bf D85} (2012) 094034.

\bibitem{SS2012} 
W. Sch\"afer and A. Szczurek,
``Production of two $c \bar c$ pairs in gluon-gluon scattering
in high energy proton-proton scattering'', 
Phys. Rev. {\bf D85} (2012) 094029.

\bibitem{MS2013}
R. Maciu{\l}a and A. Szczurek,
``Production of $c \bar c c \bar c$ in double-parton scattering within
$k_t$-factorization approach: meson-meson correlations'',
Phys. Rev. {\bf D87} (2013) 074039.

\bibitem{HMS2014} 
A. van Hameren, R. Maciula and A. Szczurek,
"Single-parton scattering versus double-parton scattering
in the production of two $c \bar c$ pairs and charmed meson correlations
at the LHC",
Phys. Rev. {\bf D89} (2014) 094019.

\bibitem{HMS2015} 
A. van Hameren, R. Maciu{\l}a and A. Szczurek,
``Production of two charm quark-antiquark pairs in single-parton
scattering within the $k_{t}$-factorization approach'',
Phys. Lett. {\bf B748} (2015) 7737.

\bibitem{LHCb_jpsijpsi}
R. Aaij et al. (LHCb collaboration) \\
``Observation of $J/\psi$ pair production in $p p$ collisions at 
$\sqrt{s}$ = 7 TeV'',
Phys. Lett. {\bf B707}, 52 (2012)

\bibitem{CMS_jpsijpsi}
V. Khachatryan et al.(CMS collaboration) \\
``Measurement of prompt $J/\psi$ pair production in $p p$ collisions at
$\sqrt{s}$ = 7 TeV'',
JHEP {\bf 1409}, 094 (2014).

\bibitem{ATLAS_jpsijpsi}
``M. Aaboud et al.(ATLAS collaboration), \\
``Measurement of the prompt $J/\psi$ pair production cross section
in $p p$ collisions at $\sqrt{s}$ = 8 TeV with the ATLAS detector'',
Eur. Phys.J. {\bf C77}, 76 (2017); arXiv:1612.02950 [hep-ex].

\bibitem{Baranov:2012re}
S.~P.~Baranov, A.~M.~Snigirev, N.~P.~Zotov, A.~Szczurek and W.~Sch\"afer,
``Interparticle correlations in the production of $J/\psi$ pairs in 
proton-proton collisions'',
Phys. Rev. D \textbf{87} (2013) no.3, 034035
doi:10.1103/PhysRevD.87.034035
[arXiv:1210.1806 [hep-ph]].

\bibitem{babiarz_pseudoscalar}
I. Babiarz, R. Pasechnik, W. Sch\"afer and A. Szczurek,
``Hadroproduction of $\eta_c(1S,2S)$ in the $k_T$-factorization approach'',
JHEP {\bf 2002} (2020) 037.

\bibitem{babiarz_scalar}
I. Babiarz, R. Pasechnik, W. Sch\"afer and A. Szczurek,
``Hadroproduction of scalar $P$-wave quarkonia in the light-front 
$k_T$-factorization approach'',
JHEP {\bf 06} (2020) 101.

\bibitem{MSS2021}
R. Maciula, W. Sch\"afer and A. Szczurek,
``On the mechanism of $T_{4c}(6900)$ tetraquark production'',
Phys. Lett. {\bf B812} (2021) 136010.

\end{thebibliography}
\end{document}